\documentclass[twocolumn, letterpaper]{emulateapj}
\usepackage{natbib}

\usepackage{epsfig}

\newcommand{\avg}[1]{\langle{#1}\rangle}

\newcommand{\ltsima}{$\; \buildrel < \over \sim \;$}
\newcommand{\lsim}{\lower.5ex\hbox{\ltsima}}
\newcommand{\gtsima}{$\; \buildrel > \over \sim \;$}
\newcommand{\gsim}{\lower.5ex\hbox{\gtsima}}

\def\gtrsim{\mathrel{\hbox{\rlap{\hbox{\lower4pt\hbox{$\sim$}}}\hbox{$>$}}}}

\def\lesssim{\mathrel{\hbox{\rlap{\hbox{\lower4pt\hbox{$\sim$}}}\hbox{$<$}}}}

\def\Cl{$C^{21}_l$}
\def\ClX{$C^{21}_l(X)$}
\def\bCl{$\bar{C}^{21}_l$}

\begin{document}
\title{Constraining Primordial Non-Gaussianities from the WMAP2
2-1 Cumulant Correlator Power Spectrum}
 
\author{ Gang Chen\altaffilmark{1} and Istv\'an Szapudi\altaffilmark{1} }
 
\altaffiltext{1}{Institute for Astronomy, University of Hawaii,
2680 Woodlawn Dr, Honolulu, HI 96822, USA}

\begin{abstract}

We measure the 2-1 cumulant correlator power spectrum \Cl, a degenerate
bispectrum, from the second data release of the Wilkinson
Microwave Anisotropy Probe (WMAP). Our high resolution measurements with SpICE
span a large configuration space ($\simeq 168\times999$) corresponding
to the possible cross-correlations of the maps recorded
by the different differencing assemblies.
We present a novel method to recover 
the eigenmodes of the correspondingly large 
Monte Carlo covariance matrix.
We examine its eigenvalue spectrum and use random matrix theory
to show that the off diagonal  terms are dominated by noise. 
We minimize the  $\chi^2$ to obtain constraints for the 
non-linear coupling parameter $f_{NL} = 22 \pm 52\, (1\sigma)$.

\end{abstract}


\section{Introduction}

Quantifying the non-Gaussianity in the cosmic microwave background (CMB)
puts constraints on inflationary models and possibly
identifies non-linear effects from large scale structure. 
In addition, systematics
and foregrounds might also produce 
non-Gaussian signatures; this possibly weakens 
constraints on the primordial and secondary non-Gaussianities.

A natural phenomenological parametrization of
non-Gaussian models is in terms of the
perturbative non-linear coupling in the primordial curvature 
perturbation \citep{KomatsuSpergel2001}:
\begin{equation}
  \label{eq:modelreal}
  \Phi({\mathbf x})
 =\Phi_L({\mathbf x})
 +f_{NL}\left(
              \Phi^2_{L}({\mathbf x})-
	      \left<\Phi^2_{L}({\mathbf x})\right>
        \right),
\end{equation}
where $\Phi_L({\mathbf x})$ denotes the linear Gaussian 
part of the Bardeen curvature and $f_{NL}$ is the 
non-linear coupling parameter. 
The resulting leading-order non-Gaussianity is at the
three-point level. Thus, the three point correlation function
\cite[e.g.,][and references therein]{ChenSzapudi2005}
or its spherical harmonic transform, the bispectrum, directly
estimate the leading order effect.

The bispectrum has been used extensively for studying non-Gaussianity
\citep{KomatsuEtal2005, CreminelliEtal2006, MedeirosContaldi2006,
LiguoriEtal2006, CabellaEtal2006}. In previous measurements,
the pseudo-bispectrum was used, which, like the pseudo-$C_l$'s,
ignores in detail the effects of the complicated geometry induced by 
Galactic cut and cut-out holes. Pixel space statistics, such
as the three-point correlation function, trivially deconvolve
the geometric effects, as the convolution kernel is diagonal in
pixel space. Indeed, SpICE \citep[][Spatially Inhomogenous Correlation
Estimator]{SzapudiEtal2001a}
uses this simple fact to estimate the angular power spectrum
without explicitly inverting the $M_{ll^\prime}$ kernel \citep{HivonEtal2002}.
For  the bispectrum, the convolution kernel is even more complex than
for the $C_l$'s, therefore pixel space methods 
are advantagous. We use the fact that the
SpICE algorithm can be used to calculate (deconvolved)
power spectrum of cumulant correlators $\avg{\delta T^N \delta T^M}$
\citep{SzapudiEtal1992}. 
These are degenerate $N+M$-point correlation functions, and
their power spectra correspond to integrated $N+M-1$ poly-spectra.
In this paper we focus on the 2-1 cumulant correlator
power spectrum which is directly related to bispectrum
configurations. These contain less configration information
than the full bispectrum, although retain more than the
skewness \citep{KomatsuEtal2005}. Note that in terms
of $f_{NL}$, with near optimal weighting, the statistical
power of the skewness is nearly as optimal as the full bispectrum
\citep{KomatsuEtal2003, SpergelEtal2006}. For models
with non-trivial configuration dependence, this is not the case.

We introduce the power spectrum of the 2-1 cumulant 
correlator together with the corresponding theoretical 
theoretical predictions. Statistical estimation in the 
data and simulations, and the theoretical calculations are
are described in \S3. 
In \S4 we investigate the covariance matrix of the measurements, 
the $\chi^2$ analysis of $f_{NL}$.
We summarize and discuss the work in \S5.

\section{power spectrum of 2-1 cumulant correlator}

For the temperature fluctuation field 
$T(\hat{\bf n})$ in CMB, the 2-1 cumulant correlator
is simply expressed as $\left<T(\hat{\bf n})^2T(\hat{\bf m})\right>$,
where the ensemble average can be replaced by spatial (angular)
average due to the assumed rotational invariance and ergodicity 
of the universe. 
Its Fourier or spherical harmonic transform 
corresponds to a set of summed (integrated) bispectrum
configurations \citep{Cooray2001}:
\begin{equation}
C_l^{21} =  \sum_{l_1 l_2} B_{l_1 l_2 l} W_{l_1}W_{l_2}W_{l} \left(
                         \begin{array}{ccc}
       l_1 & l_2  & l  \\
         0 & 0 & 0
                         \end{array}
                   \right)
\sqrt{
        (2 l_1+1) (2 l_2+1) \over 4\pi (2 l+1) } \, .
\label{eqn:finalform}
\end{equation}
Here $B_{l_1 l_2 l}$ is the bispectrum and $W_l$ is the multiple of the
pixel window function and beam function of the CMB map. 

The above equation
can be used to turn a theoretical prediction for the full bispectrum
into a prediction for the 2-1 cumulant correlator power spectrum.
We follow closely the method
described in \cite{KomatsuSpergel2001} to predict the full bispectrum
using our modified version of CAMB\footnote{http://camb.info/}.
Then we sum the above Equation~\ref{eqn:finalform} in $l_1$ and
$l_2$ up to $l_i=2000$ where $W_l \simeq 0.001$. 
Because of the linear dependency of the bispectrum on  $f_{NL}$,
we perform the calculation  with $f_{NL}=1$.

Due to the similarity of cumulant correlators to two-point
correlation functions, the analogous technique can be used
for them as for measuring $C_l$. We use cross correlations 
of a triplet of maps for each measurement to avoid 
the uncertainties in the noise bias.
The SpICE \citep{SzapudiEtal2001a} algorithm 
uses harmonic transforms to calculate fast correlation
functions of two maps, and Legendre integration
\citep{SzapudiEtal2001b} to obtain the final power spectrum.
To obtain 2-1 cumulant correlators, the first two maps have to
be multiplied first, as it is described next in detail.
The final power spectrum can be directly compared with
the theoretical prediction, as it is fully corrected for
the complicated pixel geometry of the underlying maps.
Our measurement is the first such bispectrum measurement, where
the geometry is accurately taken into account.

\section{Measurements and Predictions}

We use WMAP three year coadded foreground reduced sky maps 
\citep[][hereafter WMAP2]{JarosikEtal2006}\footnote{http://lambda.gsfc.nasa.gov/} 
with the resolution  of $n_{side}=512$. 
There are 8 differencing assemblies differencing assemblies (DAs) for the Q, V, and W bands. 
We calculated the cross correlations 
among these 8 maps and there are $(8\times7)/2\times6 = 168$ 
combinations. The factor of $1/2$ is explained by the invariance
of the 2-1 cumulant correlator under the exchange of the first
two maps. We index the triplet of maps used in the cross-correlations
by $X$  and denote the power spectrum by \ClX. First we prepare
a ``square temperature'' map by multiplying pixel by pixel the first two
maps in a triplet. The resulting map is cross-correlated by the third
map with SpICE. We use the more conservative Kp0 mask since non-Gaussianities
are expected to be more sensitive to foregrounds than the power spectrum.
It takes about 330 minutes to calculate 
all 168 spectra  on a 2 GHz class CPU
for the WMAP data (or for one set of simulations).
Figure~1 displays typical measurements  
together with theoretical predictions and simulation results
for the triplet X=(Q1,Q2,V1).

\begin{figure}[htb]
\epsscale{1.}
\plotone{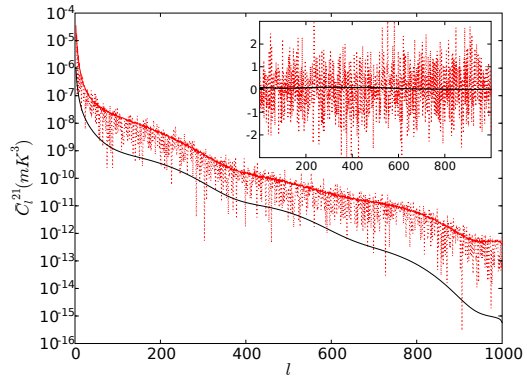}
\caption{
A typical results 
of the 2-1 cumulant power spectrum normalized to
the triplet $X = (Q1,Q2,V1)$.
The lower solid curve is the theoretical
prediction for $f_{NL}=-100$;
the upper solid line
is the standard deviation from 200 simulations;
the dotted line corresponds to the absolute values from WMAP2.
\textit{Inset}: the data (without absolute value) and theory
on a linear scale renormalized by the standard deviation measured
from the simulations.
}
\label{fig:C2l}
\end{figure}

Qualitatively, there is no obvious sign of non-Gaussianities. To 
obtain accurate constraints, we estimate covariances from
a set of Monte Carlo simulations and fit the $f_{NL}$ parameter.
Gaussian simulations ($f_{NL}$=0) suffice, since the covariance
is dominated by the Gaussian noise
as long as $f_{NL}<500$ \citep{KomatsuEtal2003}, which we already
know to be true. We generated 200 simulations with
SYNFAST in HEALPix package\footnote{http://healpix.jpl.nasa.gov/}. 
We used the WMAP first year (WMAP1) power spectrum  available from the
Lambda website
\footnote{The WMAP2 power spectrum was not available at this
writing but the difference should be insignificant for our purposes.
} for input. It is the best fit $\Lambda$CDM 
model using a scale-dependent (running) 
primordial spectral index, using WMAP1,
CBI and ACBAR CMB data, 
plus the 2dF and Lyman-alpha data. 
For each simulation we generate 8 DA maps closely mimicking the data
\citep{SpergelEtal2003}. These maps represent the same realization
of the CMB but with different noise and beam for each.
Our noise model is somewhat simplistic, and uses
Gaussian realizations  with standard deviation $\sigma_0/\sqrt{N_{obs}}$,
where the effective number of observations $N_{obs}$ varies 
across the sky and for different DAs, and $\sigma_0$  
is a constant for each DA. This does not take
into account possible correlations in the noise, but at this
writing no noise maps were available for WMAP2; our measurements in WMAP1
with more realistic noise indicate that the effects of the noise correlations
are negligible on our measurement (see the last section for details). 
Finally, each simulation was analyzed exactly the same way as the data:
we performed $33,600$ measurements in $16,00$ maps. 

\section{The Covariance Matrix and $\chi^2$ Analysis}

Let us extend our notation and
label our measurements and predictions for triplet $X$ as
$C_l^{21}(X,s)$, where $s$ stands for 
the theory ($s=0$), one of the 200 simulations($s=1, ..., 200$),
or the WMAP data($s=201$). Our goal is to obtain quantitative
constraints for $f_{NL}$ using these results. In particular,
we focus on the covariance matrix (CM).

\subsection{The Covariance Matrix}

The simulations can be used to obtain 
an experimental CM the standard way.
Without any further binning, the CM is a square matrix of size
$p=168\times999 = 167832$ (999 stands for $l=2,...,1000$).
A matrix of this size can only be inverted with supercomputers.
To overcome this problem, we introduce a novel, 
generally applicable technique which speeds up the inversion
of experimental covariance matrices. We show that the 
singular value decomposition (SVD) of the CM is related to another
matrix of smaller size governed by the number of simulations.
 
Let M be the matrix where each raws and columns correspond
to configurations and simulations, respectively.
The SVD of this $m \times n$ matrix $M$ 
(the number of rows $m$ is typically larger than the number of columns $n$)
is given by $ M = U \Lambda V^T$. Here $U$ is a $m \times n$ 
column-orthonormal matrix, $\Lambda$ is a $n \times n$ 
diagonal matrix with non-negative elements, and 
$V$ is a $n \times n$ orthonormal matrix. In this notation
the CM can be expressed as $1/nMM^T$. The corresponding
SVD is $U(1/n\Lambda^2)U^T$. This can obtain more efficiently by solving
the dual problem $M^TM=V^T\Lambda^2V$. Once the eigenvalues 
and $V$ are obtained from this smaller matrix, 
$U=MV\Lambda^{-1}$ can be calculated. This procedure is much faster
than the direct calculation of the covariance matrix, as long
as $m \gg n$. Also, it is clear from the arguments, that the
CM has at most $n$ non-degenerate eigenmodes. Unfortunately,
using only a small number of noisy eigenmodes means that
we cannot fully exploit the information content of our data.
While we have performed $\chi^2$ analysis using the full
data set, a more reliable result can be obtained
by compressing the data.

\subsection{Binning}

It would be desirable to combine different measurements
of the same $l$ with inverse variance weighting
to compress our data set. Such a combination
for the angular power spectrum is nearly optimal 
\citep{FosalbaSzapudi2004}. The generalization is
a bit more complex, since the pixel and beam window
functions for a particular triplet in Equation~\ref{eqn:finalform}
cannot be decomposed, therefore one cannot simply obtain
corrected $C^{21}_l$ estimates independent of window functions.

To overcome this problem we introduce a new unbiased quantity,
\begin{equation}
\widetilde{C}_l^{21}(X,s) =C_l^{21}(X,s)\times\frac{C_l^{21}((V1,V2,Q1),0)}{C_l^{21}(X,0)}\Big|_{f_{\rm NL}=1},
\end{equation}
the cumulant correlator power spectrum normalized according to
the theory to a arbitrary triplet $X=(V1,V2,Q1)$
\footnote{Different choices of X make no difference for our 
results, so we will keep this normalization for the rest of the paper.} 
at a fiducial value $f_{NL}=1$.
In this normalization
the window factors approximately cancel, therefore we can
obtain an inverse variance weighted 
\begin{equation}
\bar{C}^{21}_l(s) = \frac{\sum_{X=1}^{168}w(X)\widetilde{C}_l^{21}(X,s)}{\sum_{X=1}^{168}w(X)},
\end{equation} 
where $w(x)$ is proportional to the inverse variance measured
in the simulations. The advantage of the inverse variance weighting
over the noisy covariance matrix is that each weight is determined
with high accuracy $1/\sqrt{200} \simeq 0.07$, while, as we will
see, modes of the matrix are significantly affected by the noise.
Therefore, although our numerical method of the previous section
allows us to handle a matrix of very large size, we opted for
the inverse variance weighted estimators 
to suppress the noise more effectively.

\subsection{Random Matrix Theory}

To assess the level of noise in our final covariance matrix,
we use some results from random matrix theory.
\citep[e.g.,][]{SenguptaMitra1999, LalouxEtal1999a}.
As before, let us assume that a CM $C$ is constructed as
$C = 1/n M M^T$, from $M$, an $m \times n$ rectangular matrix.
Let us further assume that $M$ is composed of independent,
identically distributed random variables with zero mean and unit variance.
In the limit $m \to \infty$,  $n \to \infty$ while
$Q = n/m $ is kept fixed,
the eigenvalue spectrum will tend to:
\begin{eqnarray}
\rho_C(\lambda)&=& \frac{Q}{2\pi}\frac{\sqrt{(\lambda_{max}-\lambda)(\lambda-\lambda_{min})}}{\lambda},\\
\lambda_{min}^{max}&=&1+1/Q\pm2\sqrt{1/Q},
\end{eqnarray}
with $\lambda \in [\lambda_{min}, \lambda_{max}]$. The
eigenvalue density of  $C$,
 $\rho_C(\lambda)$ is defined as
\begin{equation}
\rho_C(\lambda)=\frac{1}{m}\frac{dn(\lambda)}{d\lambda},
\end{equation}
where $n(\lambda)$ is the number of eigenvalues of $C$ less than $\lambda$.

In our numerical construction of the covariance matrix, 
if \bCl were independent of each other, we would have a random
matrix. How significant are the correlations? 
According to Figure~2, the density of eigenvalues of our CM
appears to be consistent with the random matrix theory with $Q = 999/200$, 
with possibly a slight deviation for the largest
eigenvalues. Comparing with random matrix simulations, we find that
$Q=950/200$ would produce the same effect. While we keep
 $Q = 999/200$ in our analysis, this might be a sign of
small correlations in the CM at the $5\%$ level; this
could effect our final $\chi^2$ and our final
error-bars only slightly. 

\begin{figure}[htb]
\epsscale{1.}
\plotone{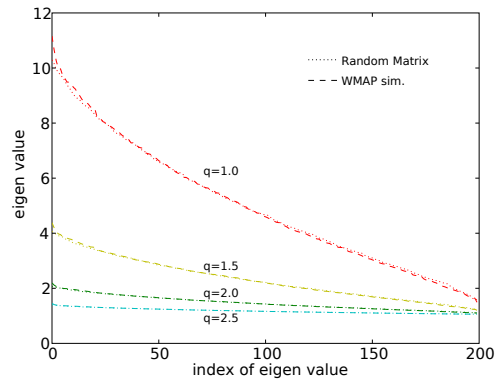}
\caption{The eigenvalue spectrum for a set of 
power mapped CMs from 200 WMAP2 simulations and 
from the 200 random matrix simulations. The original matrix
corresponds to $q=1$. To display more detail, 
we plot unbinned (sorted) eigenvalues of our CM
along with those obtained from random matrix simulations.
The spectacular agreement indicates that the random matrix
assumption is fairly accurate.
}
\label{fig:pm}
\end{figure}

To further investigate the randomness of the covariance matrix,
we apply the technique of power mapping
to our CM and random matrix simulations for comparison.
The $q$th power mapping of a matrix $C$ is defined as\citep{GuhrKalber2003}:
\begin{equation}
C_{kl}^{(q)}=\mathrm{sign}(C_{kl})|C_{kl}|^q.
\end{equation}
\cite{GuhrKalber2003} show that the eigenvalue spectral density of 
power mapped correlation matrices can detect correlations otherwise
buried in the noise. The variance of individual elements scales
as $1/n^{q/2}$, i.e. effective number of simulations increased
to  $n^q$ by power mapping.
In our case, no significant deviation
from random matrix theory appears as a result
of power mapping (see Figure~2),
which shows that the covariance is dominated by random noise.

\section{Results and Discussion}

Given that the CM of our measurement is consistent with the random matrix
assumption at the 5\% level, and that from 200 Monte Carlo simulations
7\% fluctuations are expected for each of its elements, we conclude
that it is consistent with our simulations to use a diagonal $\chi^2$
rather than weighting with the noisy off diagonal elements of CM.
We found that using diagonal $\chi^2$ is entirely robust when we
vary our binning scheme or apply no binning. On the contrary,
using the noisy eigenmodes of the CM produces
somewhat  unstable results.

Minimizing $\chi^2$ as function of $f_{NL}$ 
gives $f_{NL}=22\pm 52\, (1\sigma)$, which is the final result of
our paper.
The minimum reduced $\chi^2$ is about 1.1 for 998 degrees of freedom.
According to our discussion \S4.3 there might be an additional  $5\%$ 
uncertainty on these results because we neglected
off diagonal correlations. Our results are entirely consistent with 
\cite{SpergelEtal2006} despite that we did not weight with the theory.
The lack of optimal weighting is compensated by the
increased configuration dependence of our statistic.
Note that we did not attempt to correct for point source contamination,
but this should have a negligible effect on our results 
\citep{SpergelEtal2006}.

To test the robustness of our constraints, we divided the 200 simulations 
to two sets of 100 simulations, then repeated the
full statistical analysis
with each set. We found consistent results: 
$f_{NL}=24\pm54$ and $f_{NL}=21\pm49$, respectively.

To test the degree to which  noise correlations 
might affect our
covariance, we also repeated our analysis of the CM with the WMAP1 
simulations of \cite{ChenSzapudi2005}. where the correlated 
noise were used. The eigenvalue spectrum is virtually identical to 
what we get for WMAP2 simulations.

Finally, we repeated our analysis using the less
conservative Kp2 mask and obtained $f_{NL}=-78\pm52$.
This shows that the statistical variance of our results
is already of the same order of magnitude as the possible effect
of foreground corrections
near the galactic plane \citep{SpergelEtal2006}, therefore it
would be difficult to improve on significantly with the present data set.

We acknowledge useful discussions with Adrian Pope and Mark Neyrinck.
Some of the results in this paper have been derived using 
the HEALPix \citep{GorskiEtal2005} package.
We acknowledge the use of the Legacy Archive 
for Microwave Background Data Analysis (LAMBDA). 
Support for LAMBDA is provided by the NASA Office of Space Science.
The authors gratefully acknowledge support
by NASA through AISR NAG5-11996, NNGO06GE71G
and ATP NASA NAG5-12101 as well as by
NSF grants AST02-06243, AST-0434413  and ITR 1120201-128440.





\end{document}